\documentclass[prd,aps,preprintnumbers,fleqn,showpacs,nofootinbib,superscriptaddress]{revtex4}

\usepackage{epsfig}
\usepackage{bm}
\usepackage{amssymb}
\usepackage{amsmath}
\usepackage{color}
\usepackage{subfigure}

\begin{document}

\title{Studying Coulomb correction at EIC and EicC}

\author{Ze-hao Sun}
 \affiliation{\normalsize\it Key Laboratory of
Particle Physics and Particle Irradiation (MOE),Institute of
Frontier and Interdisciplinary Science, Shandong University,
(QingDao), Shandong 266237, China }

\author{Du-xin Zheng}
 \affiliation{\normalsize\it Key Laboratory of
Particle Physics and Particle Irradiation (MOE),Institute of
Frontier and Interdisciplinary Science, Shandong University,
(QingDao), Shandong 266237, China }

\author{Jian~Zhou}
 \affiliation{\normalsize\it Key Laboratory of
Particle Physics and Particle Irradiation (MOE),Institute of
Frontier and Interdisciplinary Science, Shandong University,
(QingDao), Shandong 266237, China }

\author{Ya-jin Zhou}
\affiliation{\normalsize\it Key Laboratory of Particle Physics and
Particle Irradiation (MOE),Institute of Frontier and
Interdisciplinary Science, Shandong University, (QingDao), Shandong
266237, China }

\begin{abstract}
We  study the gauge link contribution to the dipole type transverse
momentum dependent distributions of coherent photons, which is conventionally referred to as the Coulomb correction.
  We further propose to search for the evidence of the Coulomb correction
  in the Bethe-Heitler process in eA collisions at EIC and EicC.
\end{abstract}

\maketitle

\section{Introduction}
The study of QED processes in a strong Coulomb field has a long history pioneered by the Bethe and Maximon's seminal work~\cite{Bethe:1954zz}, in which the Furry-Sommerfeld-Manue wave functions were used to calculate
Coulomb corrections(CC) to  the pair production and bremsstrahlung cross section.
For a comprehensive review on this topic, we refer readers to the reference~\cite{Mangiarotti:2017pag}.
It later received renewed interest in the heavy ion physics community around the time
 when physics operation began at  Relativistic Heavy Ion Collider(RHIC). A lot of efforts have been made to
 compute pure electromagnetic  lepton pair production in Ultra-Peripheral heavy ion Collisions(UPC) to all orders
 in $Z\alpha $ where $Z$ is the nuclear charge number.
 The summation of multiple photon re-scattering  can be achieved  by either
 making the systematical Eikonal approximation formulated in the impact parameter
 space~\cite{Jackiw:1991ck,Ivanov:1998ka,Eichmann:1998eh,Tuchin:2009sg}
 or solving the Dirac equation in the presence of a strong Coulomb field~\cite{Segev:1997yz,Baltz:1998zb,Baltz:2001dp,Baltz:2003dy}.
 The agreement  between these
  more modern methods and the original Bethe-Maximon's results was confirmed in Ref.~\cite{Lee:1999eya}.

The total cross section of lepton pair production in UPCs is predicated to be
reduced by the Coulomb correction.
 However, there is no clear evidence of the Coulomb correction observed in heavy ion collisions so far~\cite{Baltz:2007gs,Baltz:2010mc}.
 The fact that experimental
  data   is well described by the lowest order QED
calculation~\cite{Baltz:2007kq,Klein:2018cjh,Aaboud:2018eph,Adam:2018tdm,Adam:2019mby,ATLAS:2019vxg,Lehner:2019amb,Klein:2018fmp,Zha:2018tlq}
 leaves no much room for any higher order QED effect.
On the other hand, as the experimental observation of the Coulomb correction crucially
depends on the overall normalization of
 the total cross section that suffers from various uncertainties(reliable Coulomb dissociation estimations,
 luminosity of heavy ion beams, and faithful reproduction of experimental momenta cuts, etc.), no definitive conclusion can be drawn
 at this stage.

In this work, we  study the Coulomb correction to the Bethe-Heitler(BH) process in
 eA collisions. The deviation from the single photon exchange can be experimentally checked
 by comparing with the cross section of the BH process in ep collisions. The precise determination
 of the absolute normalization is thus not required for searching the evidence of the Coulomb correction in this process.
The distribution of  the total transverse momentum
 of the scattered electron and the emitted photon is found to be very sensitive to the Coulomb correction.
It should be feasible to test our predications  at the future Electron Ion Collider(EIC) in US and the Electron Ion Collider in China(EicC).
 We notice that this subject has been addressed in
 some earlier publications~\cite{Lee:2004ina,Olsen:2003mj,Sandrock:2018ivj}.
 The present work differs from the previous studies in two aspects: 1) The problem is re-formulated
 in the framework of transverse momentum dependent factorization~\cite{Collins:1981uk}, based on which the multiple photon re-scattering
 effect is naturally  incorporated into the gauge link. In addition, the Sudakov effect arising from
 soft photon radiation can be easily took into account in our calculation. 2) The Coulomb correction to the
  linear polarization of coherent photons~\cite{Li:2019yzy,Li:2019sin} is included.  We investigate how the polarization dependent observable
 is affected by the Coulomb correction as well.

The organization  of this paper is as follows.  We  first give the matrix element definition for the dipole type
 photon TMDs in which the initial and final state multiple photon scattering in the BH process is encoded
 in a close loop gauge link. We compute the expectation value of the photon TMD
 matrix element in a boosted Coulomb potential and obtained a close form for the case of point-like charged particle.
 For an extended charge source, the photon TMDs have to be calculated numerically.
 The resulting photon TMDs   clearly deviates from the widely used
  Weizs\"{a}cker-Williams(WW) distribution when $Z$ is large. In Sec. III,  we compute the differential cross section of the
  BH process at EIC and EicC energies using the derived photon TMDs as the input.  The Coulomb correction is signaled by the ratio
  of the cross sections in eA to that in ep collisions. Furthermore,  we show that
  the  $\cos 2\phi$ azimuthal modulation  induced by the linearly polarized photons is slightly enhanced
  due to the Coulomb correction. The paper is summarized in Sec. IV.

\section{The Coulomb correction to the photon TMDs}
 In the Bethe-Heitler process, the incoming  electron multiple rescattering off the boosted Coulomb potential
  can occur either before emitting a photon or after a photon being radiated.
  At  low virtuality, the exchanged photons  coherently couple with charged heavy ion as a whole.
The multiple  coherent  Coulomb scattering  is much more pronounced
 in eA collisions than that in ep collisions, because
 the flux of coherent photons is enhanced by the factor $Z^2$.
 If the calculation is carried out in TMD factorization, the cross section
 can be expressed as the convolution of the hard part and photon TMD distributions.
 The imaginary phase accumulated from the final and initial state interactions is summarized into a close loop gauge link in the
 photon TMD matrix element.   In analogy to the gluon TMD distributions~\cite{Mulders:2000sh},
  the formal operator   definition of photon TMDs is given by,
\begin{eqnarray}
\int \frac{dy^- d^2y_\perp}{P^+(2\pi)^3} e^{ik \cdot y} \langle P |
F_{+\perp}^\mu(0)U^\dag(0_\perp) U(y_\perp) F_{+\perp}^\nu(y)  |P
\rangle \big|_{y^+=0}= \frac{\delta_\perp^{\mu \nu}}{2}
xf_1^\gamma(x,k_\perp^2)+ \left (\frac{k_\perp^\mu
k_\perp^\nu}{k_\perp^2}-\frac{\delta_\perp^{\mu\nu} }{2}\right )
 xh_1^{\perp \gamma}(x,k_\perp^2) , \label{gmat}
\end{eqnarray}
where the transverse tensor is commonly defined:
 $\delta_\perp^{\mu\nu}=-g^{\mu\nu}+p^\mu n^\nu+p^\nu n^\mu$ and
$k_\perp^2=\delta_\perp^{\mu\nu} k_{\perp\mu} k_{\perp\nu}$.
 Two photon TMDs, $f_1^\gamma$ and $h_1^{\perp \gamma}$, are the
unpolarized and linearly polarized photon distribution, respectively.
  $U^\dag(0_\perp) U(y_\perp)$ and the transverse gauge link which
  is not explicitly shown here form a close loop gauge link. $U(y_\perp)$  is defined as,
\begin{eqnarray}
U(y_\perp)= {\cal P} e^{ie\int_{-\infty}^{+\infty} dz^-
A^+(z^-,y_\perp) }.
\end{eqnarray}
 One should notice that the gauge link
 here plays the no role in  ensuring gauge invariance as photon does't carry charge.

As argued above, at low transverse momentum,  photons coherently
generated by the charge source inside relativistic nuclei dominate
 the distribution.  Both the unpolarized and polarized distributions of coherent photons can be computed
 with the Weizs$\ddot{a}$cker-Williams method.
 If one neglects the gauge link contribution,  the photon distributions associated
 with a boosted Coulomb potential are given by~\cite{Bertulani:1987tz,Vidovic:1992ik,Li:2019yzy,Li:2019sin},
\begin{eqnarray}
xf_{1,0}^\gamma(x,k_\perp^2)=xh_{1,0}^{\perp \gamma}(x,k_\perp^2)=\frac{Z^2
\alpha}{\pi^2} k_\perp^2 \left [
\frac{F(k_\perp^2+x^2M_p^2)}{(k_\perp^2+x^2M_p^2)}\right ]^2
\label{f1h1}
\end{eqnarray}
where  $F$ is the nuclear charge form factor, and $M_p$ is proton mass.
 The subscript "$0$" denotes the WW photon  distributions.
 In the small $x$ limit, two photon distributions $xf_{1,0}^\gamma$ and $xh_{1,0}^{\perp \gamma}$
 become identical~\cite{Li:2019yzy,Li:2019sin}.

The main purpose of this work is to investigate how the photon distributions are affected
 by the gauge link.  To this end, we first express
 the gauge  potential as,
\begin{eqnarray}
{\cal V}(y_\perp) \equiv e\int_{-\infty}^{+\infty} dz^-
A^+(z^-,y_\perp)= \frac{\alpha Z}{\pi} \int d^2q_\perp e^{-iy_\perp
\cdot q_\perp} \frac{F(q_\perp^2)}{q_\perp^2+\delta^2}
\end{eqnarray}
where a photon mass $\delta$ is introduced for regulating the infrared divergence.
 The  strength of the field appears in the photon TMD matrix element takes the similar form,
\begin{eqnarray}
{\cal F}^\mu(x,y_\perp) \equiv \int_{-\infty}^{+\infty} dy^-
e^{ixP^+y^-}F_{+\perp}^\mu(y^-,y_\perp)=\frac{Z e}{4\pi^2} \int
d^2q_\perp e^{-iy_\perp \cdot q_\perp}(iq_\perp^\mu)
\frac{F(q_\perp^2+x^2M_p^2)}{q_\perp^2+x^2M_p^2}
\end{eqnarray}
where $x$ is the longitudinal momentum fraction carried by photon.
The full expression of the photon TMD distributions
 incorporating the Coulomb correction(the gauge link contribution)
 is then given by,
\begin{eqnarray}
 \int \frac{ d^2 y_\perp d^2y_\perp'}{4\pi^3} \ e^{-ik_\perp \cdot (y_\perp-y_\perp')}
  {\cal F}^{\nu}(x,y_\perp){\cal F}^{*\mu} (x,y_\perp') e^{i\left [{\cal V}(y_\perp)-{\cal V}(y_\perp')\right ]}
=\frac{\delta_\perp^{\mu \nu}}{2}x f_1^\gamma(x,k_\perp^2)+ \left
(\frac{k_\perp^\mu
k_\perp^\nu}{k_\perp^2}-\frac{\delta_\perp^{\mu\nu} }{2}\right )
 xh_1^{\perp \gamma}(x,k_\perp^2)
 \label{tmd}
\end{eqnarray}
For a point-like charged particle, the close form solution of the above integration exists.
By setting $F(q_\perp^2)=1$ and $F(q_\perp^2+x^2M_p^2)=1$, one readily obtains,
\begin{eqnarray}
{\cal V}(y_\perp)  &=& 2 Z\alpha \lim_{\delta\rightarrow 0} K_0(|y_\perp|\delta)\approx
  Z\alpha \left (-2\gamma_E+\ln \frac{4}{ y_\perp^2 \delta^2} \right )
\\
{\cal F}^\mu(x,y_\perp)&= &\frac{Z e}{2\pi}\frac{y_\perp^\mu}{|y_\perp|} xM_p K_1(|y_\perp|xM_p)
\end{eqnarray}
Inserting these results into Eq.\ref{tmd},
\begin{eqnarray}
  &&  Z^2 \alpha\int \frac{d^2y_\perp d^2y_\perp'}{4\pi^4} e^{-i(y_\perp -y_\perp') \cdot k_\perp}
 \frac{y_\perp^\nu y_\perp'^\mu}{|y_\perp||y_\perp'|}x^2M_p^2
 K_1(|y_\perp|x M_p)K_1(|y_\perp'|x M_p)\left(\frac{y_\perp'^2}{y_\perp^2}\right )^{iZ\alpha}
\nonumber \\
 &=&\frac{\delta_\perp^{\mu \nu}}{2} xf_1^\gamma(x,k_\perp^2)+ \left
(\frac{k_\perp^\mu
k_\perp^\nu}{k_\perp^2}-\frac{\delta_\perp^{\mu\nu} }{2}\right )
 xh_1^{\perp \gamma}(x,k_\perp^2)
\end{eqnarray}
One notices that the infrared cutoff scale $\delta$ dependence  now drops out.
 Carrying out the integration over $y_\perp$ and $y_\perp'$, we arrive at,
\begin{eqnarray}
 &&\!\!\!\!\!\!\!\!\!xf_1^\gamma(x,k_\perp^2)  =  xh_1^{\perp \gamma}(x,k_\perp^2)
   \nonumber \\
   &=&\frac{Z^4 \alpha^3(1+Z^2 \alpha^2)k_\perp^2 }{M_p^4 x^4}
   \/_2 F_1 \! \! \left [1\!-\!iZ\alpha, 2\!-\!iZ\alpha,2, \frac{-k_\perp^2}{M_p^2x^2} \right ]
  \/_2 F_1 \! \! \left [1\!+\!iZ\alpha, 2\!+\!iZ\alpha,2, \frac{-k_\perp^2}{M_p^2x^2}  \right ]\!
  \left ( \frac{2}{e^{Z\alpha \pi}-e^{-Z\alpha \pi}}\right )^2
  \label{11}
 \end{eqnarray}
 where $\/_2 F_1$ is the hypergeometric function. The unpolarized
  and the linearly polarized photon TMDs remain the same after
 taking into account gauge link contribution. The similar relations between  the dipole type gluon TMDs have been established
 in earlier work~\cite{Metz:2011wb,Zhou:2013gsa,Boer:2015pni} for both cases of unpolarized target
 and transversely  polarized target.
 In the limit $Z\rightarrow 1$, the above
 result is reduced to,
\begin{eqnarray}
xf_1^\gamma(x,k_\perp^2)  =  xh_1^{\perp \gamma}(x,k_\perp^2)
   \approx \frac{Z^2 \alpha}{\pi^2}
  \frac{k_\perp^2}{(k_\perp^2+M_p^2x^2)^2}
 \end{eqnarray}
which recovers Eq.\ref{f1h1} as it should. Furthermore, when
 $k_\perp^2 \gg x^2 M_p^2 $, the photon TMDs are simplified as,
 \begin{eqnarray}
xf_1^\gamma(x,k_\perp^2)  =  xh_1^{\perp \gamma}(x,k_\perp^2)
   \approx \frac{Z^2 \alpha}{\pi^2}  \frac{1}{k_\perp^2}
 \end{eqnarray}
This indicates that the  gauge link contribution, i.e. the Coulomb correction, to the photon TMDs is vanishing with increasing
 transverse momentum for the case of point like particle.
 The photon distributions are altered by the multiple re-scattering effect only
 in the low transverse momentum region. However, this is no longer true for an extended charge source as shown below.
\begin{figure}[htpb]
\includegraphics[angle=0,scale=0.9]{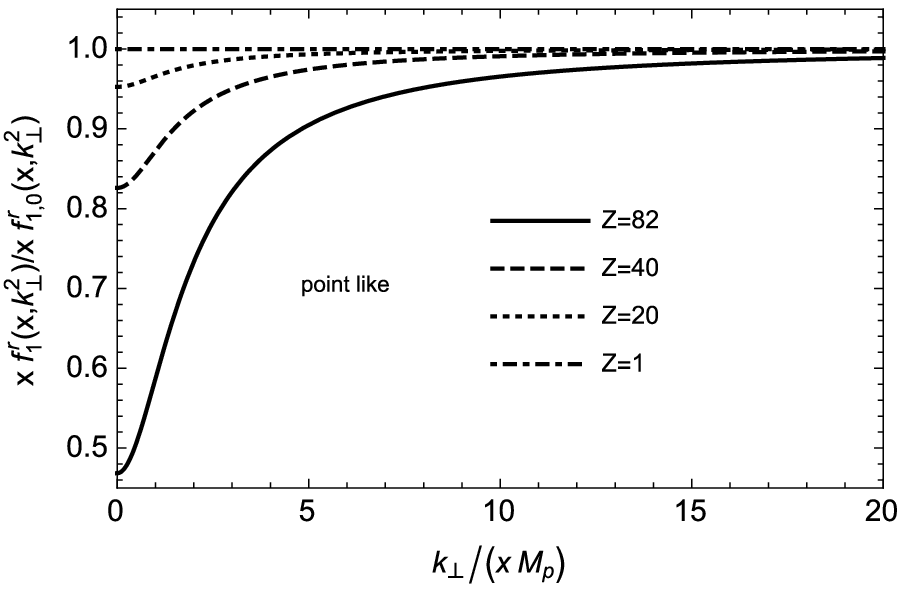}
\includegraphics[angle=0,scale=0.923]{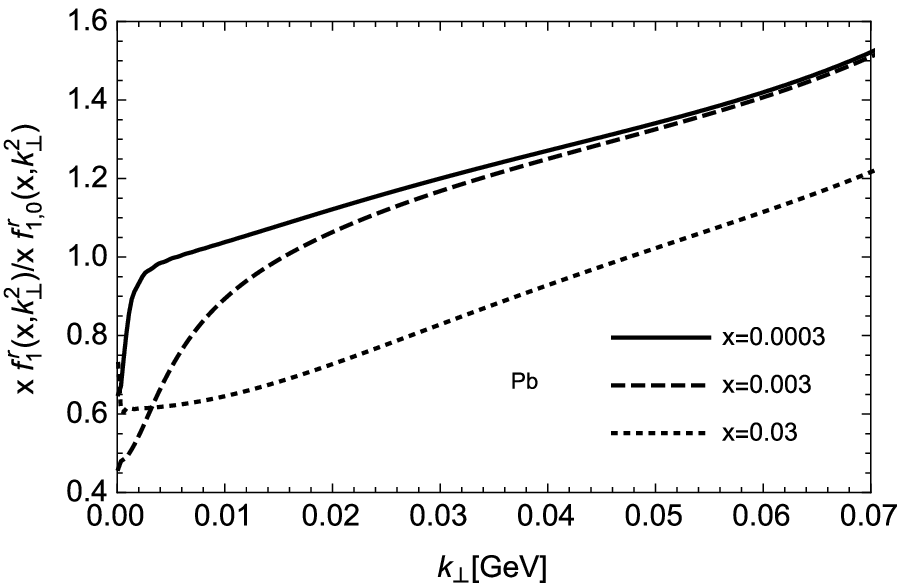}
\caption{The ratio $R=f_1^\gamma/f_{1,0}^\gamma$ is plotted as the function of  $\frac{k_\perp}{xM_p}$
for a point like charged particle(left panel).
 The same ratio is plotted as the function of  $k_\perp$ for a Pb target
at different $x$(right panel).}
\label{fig1}
\end{figure}

According to Eq.~\ref{11},  $k_\perp^2 xf_1^\gamma(x,k_\perp^2) $ is a function of
 the single variable $\frac{|k_\perp|}{xM_p}$ rather than of two variables $|k_\perp|$ and $x$.
 In Fig.~\ref{fig1}, we plot the ratio $R=f_1^\gamma/f_{1,0}^\gamma$ as the function of  $\frac{|k_\perp|}{xM_p}$
  for a point-like particle with the various choices of $Z$. One sees that
  the photon TMD is significantly reduced by
 the Coulomb correction  at the low value of $\frac{|k_\perp|}{xM_p}$.
  In the case of an extended particle, the photon distribution is no longer the
  function of the single variable $\frac{|k_\perp|}{xM_p}$.
    The ratio as the function of
 $k_\perp$ at different $x$ for a Pb target is displayed in Fig.~\ref{fig1}(right).
 In our numerical estimation, the nuclear charge  form factor is taken from the STARlight MC generator~\cite{Klein:2016yzr},
\begin{eqnarray}
F(|\vec k|)=\frac{4\pi \rho^0}{|\vec k|^3 A}\left [ \sin(|\vec
k|R_A)-|\vec k|R_A \cos(|\vec k|R_A)\right ]\frac{1}{a^2 \vec k^2+1}
\label{Af}
\end{eqnarray}
where $R_A=1.1 A^{1/3}$fm, and $a=0.7$fm.
This parametrization  is very close to the Woods-Saxon distribution.
 Our numerical results  demonstrate that the photon distribution of the charged heavy ion is also
 suppressed at low $k_\perp$ due to multiple Coulomb re-scattering.
 However, in a sharp contrast with the point like particle case,
 one notices that the ratio exceeds 1 at relatively large $k_\perp$.

  It is also interesting to investigate how the integrated photon
 distribution is modified by the Coulomb phase. The integration over $k_\perp$  has to be carried out
 with extreme caution~\cite{Lee:1999eya}. For a point like particle,
 the difference between the integrated dipole type photon distribution(with the Coulomb correction)
 and the integrated WW photon distribution(without the Coulomb correction) is given by,
 \begin{eqnarray}
\int d^2 k_\perp \left [x f_1^\gamma(x,k_\perp^2) -xf_{1,0}^\gamma(x,k_\perp^2) \right ]
=-\frac{2 Z^2 \alpha}{\pi}f(Z\alpha)
 \end{eqnarray}
 where $f(Z\alpha)\equiv Re \psi(1+iZ\alpha)+\gamma_E  $
 with $ \psi(x)=d \ln \Gamma(x)/dx$ is just the well known universal function derived in the
 Bethe-Maximon theory~\cite{Bethe:1954zz}. We also numerically test this relation
  and confirm its validation.
This  is a quite puzzling result in the sense that photon PDF seems to be
process dependent.\footnote{ If the differential cross section of forward dilepton production in ultra-peripheral heavy ion collisions
 is computed in the  modern small $x$ formalism, a photon quadruple amplitude shows up.
 In the correlation limit, the four point function collapses into two point function.
 For the Abelian case, Wilson lines and conjugate Wilson lines in the two point function
 completely cancel  out. As a consequence, the gauge link in the corresponding photon TMD is absent.
 Therefore, photon PDF in this process is free from the Coulomb correction and thus different from the
 one under consideration. }
We will thoroughly explore this issue in a future publication.

\

\section{Observalbes}
The  photon TMDs with a close loop gauge link can be probed in the
 Bethe-Heitler process,
\begin{eqnarray}
e( \bar P)+\gamma(x P+k_\perp) \rightarrow
\gamma(p_1)+e(p_2),
\end{eqnarray}
where $x P+k_\perp$ is understood as the total momentum transfer via
multiple  photon exchange. We focus on a specific kinematical region, the so-called correlation limit
where the total transverse momentum $k_\perp=p_{1\perp}+p_{2\perp}$ of the final state produced
particles($\gamma+e$) is much smaller than $P_\perp=\frac{p_{1\perp}-p_{2\perp}}{2} \approx p_{1\perp} \approx - p_{2\perp}$. In such a region, the
 calculation of the cross section can be formulated either in the CGC framework or in the TMD formalism.
 The equivalence of the two approaches has been verified
 for the gluon initiated bremsstrahlung  process~\cite{Metz:2011wb,Boer:2017xpy}.
 Obviously, all the analysis can be extended to the corresponding QED process.
  Since there are two well separated scales in the correlation limit,  large logarithm terms
   arise from unobserved soft photon radiations show up in higher order QED calculations.
It is conventional to express
 the differential cross section  in the impact parameter space to facilitate resumming these large logarithms,
 \begin{eqnarray}
\frac{d\sigma}{dP.S}=H_{\text {Born}} \int
\frac{d^2 r_\perp}{(2\pi)^2} e^{i r_\perp \cdot q_\perp}
 e^{- \frac{\alpha_e}{2\pi} {\rm ln}^2 \frac{P_\perp^2}{\mu_r^2}}
\int d^2 k_\perp
e^{i r_\perp \cdot k_\perp} xf_1^\gamma(x,k_\perp^2)
 \end{eqnarray}
with $\mu_r=2 e^{-\gamma_E}/|r_\perp|$.
Here the Sudakov factor $ e^{- \frac{\alpha_e}{2\pi}{\rm ln}^2 \frac{P_\perp^2}{\mu_r^2}} $
   takes care of all order soft photon radiation effect up to the double leading logarithm accuracy.
The phase space factor is defined as $d P.S=d y_\gamma  d^2P_\perp
d^2 q_\perp$, where $ y_\gamma$ is the  rapidity of the
 emitted photon. The hard coefficient is given by,
\begin{eqnarray}
H_{\text{Born}}=2 \alpha_{e}^2   z^2
  \frac{1+(1-z)^2}{P_{\perp}^4}
\end{eqnarray}
 where $z$ is  the
longitudinal momentum fraction of the incoming electron carried by the  final state photon.
In the above formula, the electron mass has been neglected. This is a very good approximation at EIC and EicC
energies.

\begin{figure}[htpb]
\includegraphics[angle=0,scale=0.9]{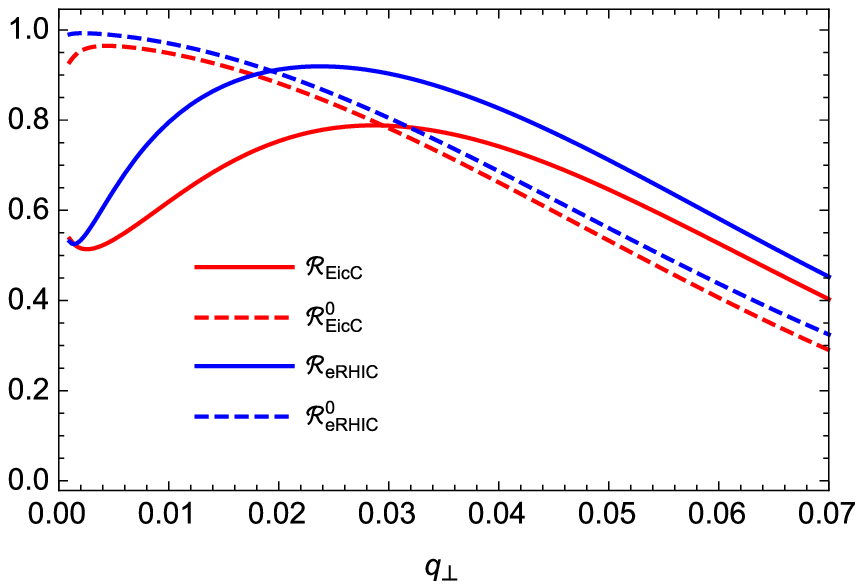}
\includegraphics[angle=0,scale=0.9]{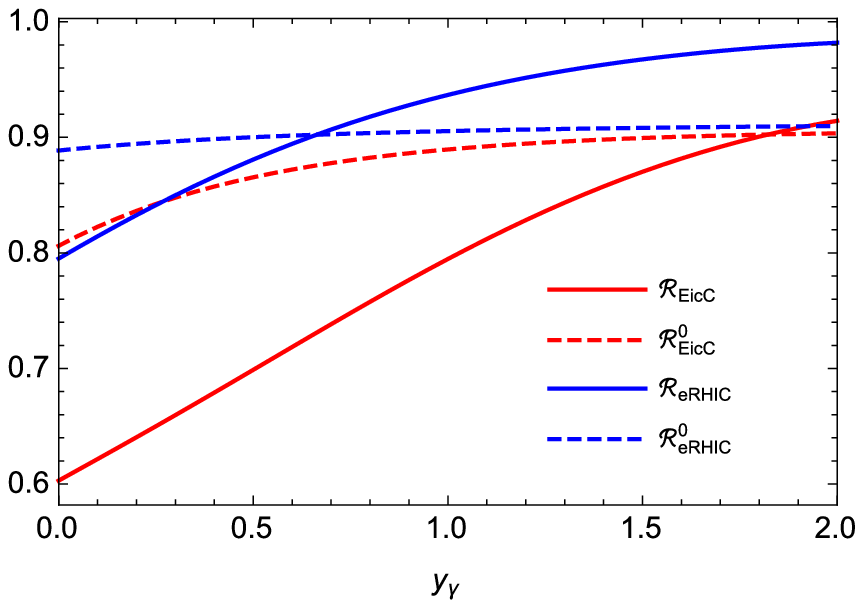}
\caption{ (color online) The ratio  ${\cal R}$ and  ${\cal R}^0$  as the function of
$q_\perp$(left panel) and $y_\gamma$(right panel)  for a Pb target
at EicC and EIC.
  $P_\perp$ is integrated over the regions [300 MeV, 400 MeV] for EicC and [1.5 GeV, 2 GeV] for EIC.
 In the left plot, the emited photon rapidity $y_\gamma$ is integrated over [0.5, 1].
 In the right plot, the total transverse momentum
  $q_\perp$ is fixed to be 20MeV. }
  \label{fig2}
\end{figure}
The Fig.~\ref{fig2} displays the two ratios $ {\cal R}(q_\perp) $ and $ {\cal R}^0(q_\perp)$ defined as follows,
\begin{eqnarray}
 {\cal R}(q_\perp) =\frac{d\sigma_{eA}}{Z^2 d\sigma_{ep}} \  \   \  \ \
    \  \ {\cal R}^0(q_\perp)=\frac{d\sigma_{eA}^0 }{Z^2d\sigma_{ep}^0}
\end{eqnarray}
where $d\sigma_{ep}$ and $d\sigma_{eA}$ are the exact cross sections of the BH process in ep and eA
 scatterings respectively. For a comparison, the cross sections $d\sigma_{ep}^0$ and $d\sigma_{eA}^0$
  are computed using the conventional equivalent photon approximation(see Eq.\ref{f1h1}).
  The parametrization $F(|\vec k|)=1/(1+\frac{\vec k^2}{Q_0^2})^2$ with $Q_0^2=0.71 \text{ GeV}^2$
   for the proton charge form factor is used for determining  the photon distribution of proton.
  Note that the Coulomb correction to photon distribution of proton is negligible.
  Therefore, we simply use  the WW photon distribution to calculate the cross section in ep collisions.
  The proton magnetic moment contribution to the BH cross
    section at low $q_\perp$ can be neglected.
    The rapidity $y_\gamma$ in the Fig.~\ref{fig2}(right) is defined in the lab frame where
   electron beam and heavy ion beam energies are 18 GeV and 100 GeV for EIC respectively,
 while they are 3.5 GeV and 8 GeV for EicC respectively.
   From Fig.~\ref{fig2}, one sees that the ratios $\cal R$  and ${\cal R}_0$ are rather different
  in the most kinematical regions at EIC and EicC energies.
     If the experimentally measured ratios deviate from these dashed lines(${\cal R}_0$) presented in Fig.~\ref{fig2}, it
 would be a clear evidence of the Coulomb correction.

We now turn to study the impact of the Coulomb correction on the polarization
dependent observable in the BH process.  A $\cos 2\phi$ azimuthal modulation in the BH cross section is induced
 by the linearly polarized photons if a virtual photon instead of real one is emitted in the final state.
 The similar phenomena in  QCD has been studied in Ref.~\cite{Metz:2011wb,Boer:2017xpy},
 from which one can readily recover the azimuthal dependent cross section in the QED case,
\begin{eqnarray}\nonumber
\frac{d\sigma}{dP.S}&=&\int
\frac{d^2 r_\perp}{(2\pi)^2} e^{i r_\perp \cdot q_\perp}
 e^{- \frac{\alpha_e}{2\pi} {\rm ln}^2 \frac{Q^2}{\mu_r^2}}
 \\&&\times \int d^2 k_\perp
e^{i r_\perp \cdot k_\perp} \Big{\{}  H_{\text{Born}}'xf_1^\gamma(x,k_\perp^2)
 + H_{\text{Born}}^{\text{cos}(2\phi)} \left [ 2(\hat{k}_{\perp} \cdot \hat{P}_{\perp})^2-1 \right ]
 x h^{\perp \gamma}_{1}(x,k_{\perp}^2)   \Big{\}}
\end{eqnarray}
where  $\hat k_\perp=k_\perp/|k_\perp|$ and $\hat
P_\perp=P_\perp/|P_\perp|$ are unit transverse vectors.  The hard parts take the form,
\begin{eqnarray}
\nonumber H_{\text{Born}}'&=& 2 \alpha_{e}^2    z^2
 \left[  \frac{1+(1-z)^2}{\left(P_{\perp}^2+(1-z)Q^2\right)^2}
 -\frac{2 Q^2 P^2_{\perp} z^2 (1-z)}{\left(P_{\perp}^2+(1-z)Q^2\right)^4} \right ]
 \\
H_{\text{Born}}^{\text{cos}(2\phi)}&=& 2 \alpha_{e}^2   z^2  \frac{-2 Q^2
P^2_{\perp} z^2 (1-z)}{\left(P_{\perp}^2+(1-z)Q^2\right)^4}
\end{eqnarray}
To avoid having to deal with a three scale problem,
we restrict to the kinematical region where $Q^2$ is of the order of $P_\perp^2$.  This happens
to be the optimal region to observe $\cos 2\phi$ azimuthal asymmetry as suggested by our numerical estimation.
Moreover, as long as $Q^2$ is sufficiently large,
  the Coulomb  multiple rescattering effect can be neglected for the lepton pair production via the virtual
  photon decay.\footnote{See the footnote 1.}

\begin{figure}[htpb]
\includegraphics[angle=0,scale=0.9]{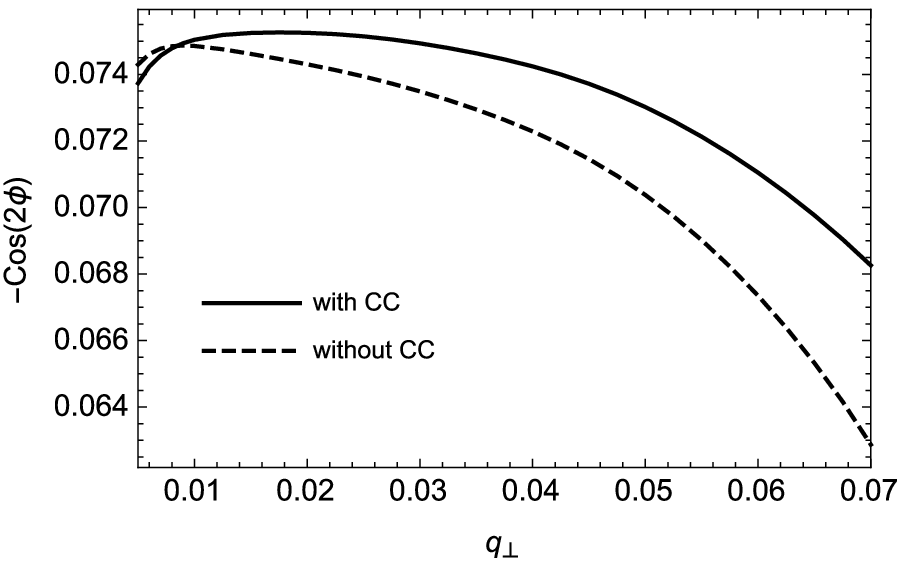}
\includegraphics[angle=0,scale=0.9]{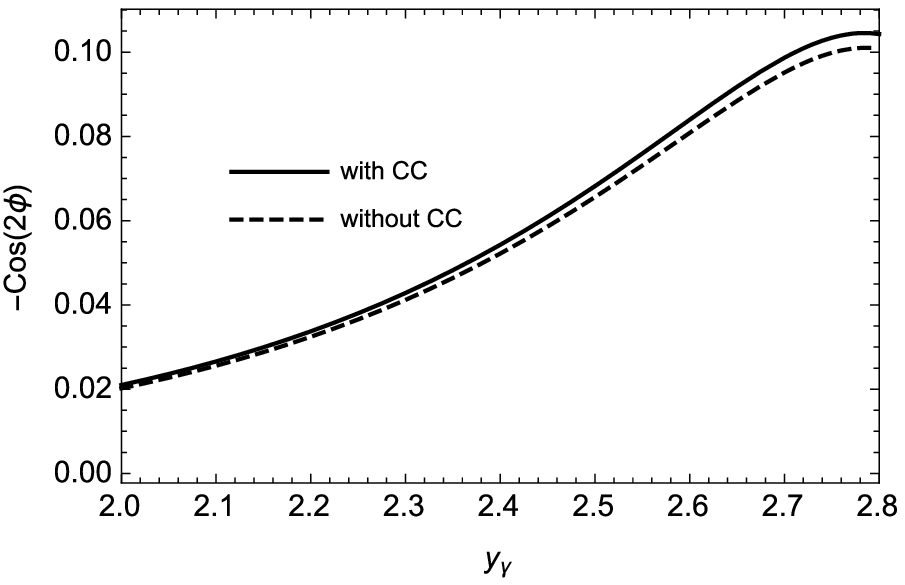}
\caption{ The azimuthal asymmetry as the function of  $q_\perp$(left panel) and $y_\gamma$(right panel)
with and without taking into account the  Coulomb corrections for a Pb target
at EIC. $Q^2$ is fixed to be  $Q^2 =4 \ \text{GeV}^2$.  The asymmetry is averaged over the $P_\perp$ region [1.5 GeV, 2 GeV].
In the left plot, the emitted photon rapidity $y_\gamma$ is integrated over the region [2, 2.8].
In the right plot, the total transverse momentum $q_\perp$ is fixed to be 50 MeV. }
\label{fig3}
\end{figure}

 We plot the azimuthal asymmetries  computed  for EIC  energy in Fig.~\ref{fig3}.
  Here the azimuthal asymmetries,  i.e. the average value of  $\cos 2\phi$ is defined as,
\begin{eqnarray}
\langle \cos(2\phi) \rangle &=&\frac{ \int \frac{d \sigma}{d {\cal P.S.}} \cos 2\phi \ d {\cal P.S.} }
{\int \frac{d \sigma}{d {\cal P.S.}}  d {\cal P.S.}}
\label{fig5}
\end{eqnarray}
As shown in Fig.~\ref{fig3}(right), the asymmetry becomes larger with the increasing photon rapidity.
The maximal value of the asymmetry reaches roughly 10\%.
According to Eq.\ref{11},   the linearly polarized
   photon TMD and the unpolarized  photon TMD are modified by the multiple Coulomb rescattering effect
    in the same way, and thus remain identical. If  the Sudakov effect were not considered, the
    azimuthal asymmetry would not be affected by the Coulomb correction.
    However,  the azimuthal averaged cross section and $\cos 2\phi$ dependent part
 evolve with the scale $P_\perp$ following  a different pattern.
 The different initial conditions for the photon distributions would lead to the different $\cos 2\phi$ asymmetries
 at  higher scale $P_\perp$.  Fig.~\ref{fig3} displays the $\cos 2\phi$ asymmetries computed with
 the photon distributions given in  Eq.~\ref{tmd} and the WW photon distributions.   One sees that at EIC,
 the deviation caused by the Coulomb correction is visible though  tiny.  On the other hand, at EicC, the difference(not shown here) is
 completely negligible since the evolution effect is much weaker at low energy scale. Therefore,
 it appears  to be not optimistic to observe the Coulomb correction effect via
 the polarization dependent observable at neither  EIC nor EicC.

\section{Summary}
In this paper, we performed the detailed analysis of the dipole type photon TMDs
 associated with a boosted Coulomb potential.
 Our main focus is on the contribution of the close loop gauge link to photon transverse momentum distributions,
 which is conventionally  refereed to as the
 Coulomb correction in the study of strong field QED. Due to the large $Z$ enhancement, the Coulomb correction(or gauge link
 contribution) alters transverse momentum distributions of photons substantially  for a charged heavy ion target,
 as compared to the Weizs\"{a}cker-Williams photon distribution.
 The photon TMDs under consideration can be accessed in the
 BH process. Our numerical results show that it is promising to observe the Coulomb correction
 at EIC and EicC. The investigation of the Coulomb correction in the BH process will
  offer us a clean way to test the TMD formulation of initial/final state
 multiple  re-scattering effects, and would be  beneficial for
 deepening our understanding of the gauge link contribution  in QCD processes.
 Moreover, the accurate account of the Coulomb correction to the BH process
  is also important for the determination of luminosity at EIC and EicC.

\begin{acknowledgments}
 J. Zhou has been supported by the National Science Foundations of
China under Grant No.\ 11675093, and by the Thousand Talents Plan
for Young Professionals. Ya-jin Zhou has been supported by the
National Science Foundations of China under Grant No.\ 11675092.
\end{acknowledgments}

\end{document}